# సామాన్య జాగృతి-పరిష్కారం

జయంతి, సిద్ధార్థ విశ్వేశ్వర

జూలై 7, 2022

జాగరితస్థానో బహిష్ప్రజ్ఞః సప్తాఙ్గ ఏకోనవింశతిముఖః స్థూలభుగ్వైశ్వానరః ప్రథమః పాదః[9]

### సారాంశం

ఈ లేఖలో సమకాలిక వినా-ప్రతీక్ష రేఖీయ దత్తాంశసంవిధానాల (కృత్రిమాణు వస్తువుల) సంశ్రమ అధోబంధాలను పరికి-స్తాం. వాఙ్మయములో, సమకాలిక ప్రతికూలతమ శ్రమవిషమత అధోబంధాలను స్థాపించేడానికి జయంతి జాగృతి పరిష్కరం ఒక ముఖ్య పరికరం. కానీ జయంతి జాగృతి-పరిష్కార సమస్యను $n$ సంసాధకాలు $O(n)$ సంశ్రమతో పూరించే విధికల్పమును ప్రదర్శించి, తద్వారా ఆ పరికరంతో సంశ్రమ అధోబంధ నిరూపణ దుర్లభం అని ప్రదర్శిస్తాం. సంశ్రమ అధోబంధాలను నిరూపిం-చేడడానికి తోడ్పడే సామాన్య-జాగృతి-పరిష్కారం అనే కొత్త పారమితిక సమస్య శ్రేణిని నిర్వచించి, దాన్ని పరిశీలించి, శ్రేణిలోని ప్రతీ సమస్యకు సంశ్రమ అధోబంధాన్ని నిరూపిస్తాం. శ్రేణిలో కష్టతమమైన సమస్యను $n$ సంసాధకాల సంవిధానము పూరించేలంటే $\Omega(n \log n)$ సంశ్రమ ఆవశ్యకం. ఈ సిద్ధాంతాన్ని ఉపయోగించి, న్యూనీకరణలద్వారా వివిధ బహుసంసాధక వస్తు ఉపవిధులకు సంశ్రమ అధోబంధాలను నిరూపిస్తాం. ఆ పై, శ్రేణిలో కొన్ని ముఖ్యసమస్యలకు మా అధోబంధాలే ఉద్బంధాలని రచనాత్మకంగా నిరూపించి, తద్వారా మా ఆ అధోబంధాలను బలపరచేట అసాధ్యమని గమనిస్తాం.

## 1 ఉపోద్ఘాతము

వేగమైన విధికల్పాలకు వెన్నెముకలు దత్తాంశసంవిధానాలు. సమకాలికసంగణనములో CASతో కల్పించేన రేఖీయ [5] వినా-ప్రతీక్ష [4] దత్తాంశాలు సువర్ణప్రమాణాలు. అందున, ఇలాంటి **కృత్రిమాణు**దత్తాంశాల శ్రమవిషమతను నిర్ధారించేట ఒక ముఖ్యాంశం. శ్రమవిషమత ఉద్బంధాలను విధికల్ప పరిశీలనతో గ్రహించేచ్చు. శ్రమవిషమత అధోబంధాలను నిర్ధారించేటకు ఉపకరించే ఒక ప్రధాన ఆధునిక సాధనం జాగృతి-పరిష్కార సమస్య.

**జాగృతి-పరిష్కార** సంసాధక సమన్వయ కర్తవ్యమును జయంతి 1998లో ప్రవేశపెట్టారు [6]. జాగృతి-పరిష్కారం ఒక $n$ సంసాధక అసమకాలిక సంవిధానంలో నిర్వచింపబడినది. దీనిలో కేంద్రబిందువు, ఉన్న $n$ సంసాధకాలలో ఒకటి మిగతా అన్ని సంసాధకాలు **లేచి** కనీసం ఒక అడుగు వేసాయని గ్రహించేలి.

ఈ జాగృతి-పరిష్కార సమస్యను కేవలం **పఠ** (read), **లిఖ** (write), **పోల్చివ్రాత**లతో (CAS) మాత్రం పూరించే ప్రతి విధిక-ల్పములో ఏదో ఒక సంసాధకం ప్రతికూలతమ చలనలో $\Omega(\log n)$ శ్రమిస్తుందని జయంతి నిరూపించేరు. ఈ సిద్ధాంతాన్ని బట్టి న్యూనీకరణల ద్వారా వివిధ సమకాలిక-దత్తాంశ-సంవిధాన విధికల్పాల ప్రతికూలతమ ఉపవిధులకు అధోబంధాలు నిరూపించే-గలిగారు.

ఈ లేఖలో, మేము సమకాలికదత్తాంశసంవిధానాల ప్రతికూలతమ సంశ్రమవిషమతను విచారిస్తాం. ఈ విచారణలో ప్రథ-మాంశముగా, జాగృతి-పరిష్కారపు సంశ్రమను విశ్లేషించి, ఈ సమస్యను కేవలం $O(n)$ సంశ్రమతో పూరించచ్చని విధికల్పనద్వారా నిరూపిస్తాం. పరిణామతః, జాగృతి-పరిష్కార న్యూనీకరణలద్వారా దత్తాంశసంవిధానాల సంశ్రమవిషమతలకు అధోబంధాలు స్థా-పించేట దుర్లభం. ఈ తడను అధిగమించేటకు, జాగృతి-పరిష్కారమును సామాన్యపరచి, సామాన్య-జాగృతి-పరిష్కారం అనే కొత్త సమస్య శ్రేణిని ప్రవేశపెడతాం. ఈ శ్రేణిలో జయంతి నిర్వక్త జాగృతి-పరిష్కారం సులభతరమైన సమస్య, అందుచేత దన్ని **సులభ జాగృతి-పరిష్కారమని** పునర్నామకరణం చేస్తాం. జయంతి సులభ సమస్య విశ్లేషణను విస్తరించి, శ్రేణిలోని ప్రతి సమ-స్యకు అధోబంధాన్ని స్థాపిస్తాం. ఉదాహరణకు, శ్రేణిలోని కష్టతమమైన **కఠిన జాగృతి-పరిష్కార** సమస్యకు సంశ్రమ అధోబంధం

$\Omega(n \log n)$. అధోబంధ స్థాపనానంతరం, న్యూనీకరణల ద్వారా వివిధ దత్తాంశసంవిధానాలకు సంశ్రమ అధోబంధాలను నిరూ‌పిస్తాం. లేఖ తుది అంశముగా, అధోబంధాల దృఢీకరణశక్యతను విశ్లేషిస్తాం. ఈ దిశలో, ముఖ్యముగా కఠిన జాగృతి-పరిష్కరణ సంశ్రమ అధోబంధానికి సమమైన ఉద్బంధాన్ని విధికల్ప ప్రమాణముగా నిరూపిస్తాం.

## 2 యంత్రప్రతికృతి మరియు పూర్వాంశాలు

ఈ లేఖలో సంవిభక్త-స్మృతి గల అసమకాలిక బహుసంసాధకమును ప్రస్తుతిస్తాం. అట్టి సంగణక-సంవిధంలో **స**$_1$, ..., **స**$_n$ అనే $n$ సంసాధకాలు, అసమకాలికముగా (అనగా వేరువేరు వేగాలతో అడుగులేసుకుంటూ)విధికల్పమును నిర్వర్తిస్తాయి. ప్రతి సంసాధ‌కం **స**$_i$ మొద్దట్టి అడుగు వేసినప్పుడు అది **లేచిందని** (అనగా **జాగృతం** అని) గణిస్తాం. లేచిన తరువాత, ప్రతి సంసాధకము విధికల్పములో దానికివ్వబడిన ఉపవిధిని కాలబంధము అవకాశమిచ్చినప్పుడల్ల ఒక అడుగు వేస్తూ తిరుగిచ్చేదాకా నిర్వర్తిస్తుంది. సంవిధము అసమకాలికము కనక, **రిపు-కాలబంధం** ఏ సంసాధకము ఎప్పుడు మరుసటి అడుగు వెయ్యాలో నిర్దేశిస్తుంది. అన‌గా, కాలబంధం ℕలోని ప్రతి **కాలమాత్ర**కి, ఆ మాత్రలో ఏ సంసాధకము అడుగువేస్తుందో నిశ్చయిస్తుంది. కాలబంధం రిపువు కనక వివిధ సంసాధకాల అడుగుల మిశ్రమ-క్రమము ఏ విధముగానైనా ఉండచ్చే. ప్రతీ మాత్రలో సరిగ్గా ఒక సంసాధకము మాత్రం అడుగు వేస్తుంది కనక సంవిధములో ప్రతీ సంసాధక అడుగుకు ఒక **సంవిధకాలమాత్ర** అని లెక్క. అలాగే, ఆ అడుగు వేసిన సంసాధకము **స**$_i$కి ఒక **శ్రమ** పడినట్లు లెక్క. సంసాధకాల శ్రమలను కూడితే ఒచ్చే సంఖ్యని సంవిధముయొక్క **సంశ్రమ** అని ప్రస్తు‌తిస్తాం. బహుసంసాధక సంగణన పరిభాషలో, ఒక ప్రత్యేక కాలబంధానుసారం సంసాధకాలచే నిర్వహింపబడ్డ విధికల్పమును **చలన** అంటారు.

ఈ భాషలో సుతధ్యముగా ప్రసాద్ జయంతియొక్క జాగృతి-పరిష్కారాన్ని ఇలా నిర్వచించచ్చు.

**నిర్వచనం 2.1** (జాగృతి-పరిష్కారం)**.** ఒక $n$ సంసాధక సంవిధానంలో **జాగృతి-పరిష్కరం** చేసే విధికల్పమునకు మూడు గుణ‌ములుండాలి:

1. **సమాప్తి:** ప్రతి సంసాధకము ఒక సదసత్తు (అనగా సత్, లేదా అసత్ ను) తిరుగివ్వాలి.

2. **సత్యవాక్కు:** అన్ని సంసాధకాలు లేచేదాకా (ఒక అడుగు వేసేదాకా), ఏ సంసాధకము సత్తును తిరుగివ్వకూడదు.

3. **అవితండం:** అన్ని సంసాధకాలు అసత్తును తిరుగివ్వకూడదు.

ఆయన ఈ సమస్యను గురించి నిరూపించిన రెండు అధోబంధాలను ఇక్కడ పునఃస్మరిద్దాం:

**సిద్ధాంతం 2.2** (అవిచక్షణీయత [6]లో Lemma 5.2)**.** $P$ అనే గణం లోని సంసాధకాలకు జాగృతి-పరిష్కారాన్ని **వ** అనే యాదృచ్చి‌కవిధికల్పం కేవలం పఠ, లిఖి, పోల్చివ్రాతలను భరించే $V$ అనే గణం వికారులతో పూరిస్తుందని భావిద్దాం. అప్పుడు క్రింది గుణాలను పూర్తి (అనగా 1) సంభవతతోగల కాలబంధం **క** ఉండితీరాలి:

1. **క** అనే కాలబంధం కల్పాల్లో చలిస్తుంది. ప్రతి కల్పంలో తిరుగివ్వని ప్రతి సంసాధకము సరిగ్గా ఒక అడుగు వేస్తుంది.

2. **అ** $\in P \cup V$ అనే ప్రతి వస్తువుకు, ప్రతి కల్పం $k$ లో జ్ఞా$_k$(**అ**) $\subseteq P$ అనే సంసాధక ఉపగణం జోడింపబడి ఉంటుంది. ఈ గణాలను **జ్ఞాన గణాలని** సంబోధిస్తాం.

3. **స** అనే సంసాధకాన్ని ధ్రువపరుద్దాం. **క** కాలబంధ క్రమములో అన్ని సంసాధకాలు చలించినా, **క** కాలబంధ క్రమములో జ్ఞా$_k$(**స**) లోని సంసాధకాలు మాత్రము చలించినా **స** దృష్టిలో కనీసం $k$వ కల్పం వరకు అవిచక్షణీయం.

4. |జ్ఞా$_k$(**స**)| $\leq 4^k$

పై సిద్ధాంతాన్ని అనుసరించి వచ్చే అధోబంధము:

**సిద్ధాంతం 2.3** (జయంతి ప్రతికూలతము అధోబంధ సిద్ధాంతం [6]లో Theorem 6.1)**.** $P$ అనే గణం లోని సంసాధకాలకు జాగృతి-పరిష్కారాన్ని **వ** అనే యాదృచ్చివిధికల్పం కేవలం పఠ, లిఖి, పోల్చివ్రాతలతో పూరిస్తుందని భావిద్దాం. $|P| = n$ ఐన, అప్పుడు, ఏదో ఒక కాలబంధం **క** ని అనుసరించి **వ** చలించినప్పుడు, ఏదో ఒక సంసాధకము **స** $\in S$ చలనలో $\Omega(\log n)$ శ్రమిస్తుంది.

ఈ లేఖలో అధోబంధాలే కాక ఉద్బంధాలను కూడా పరిశీలిస్తాం. ఆ ఉద్బంధాలను ప్రతిష్టించే విధికల్పాలకు ఉపకరించే ఇంకొక ముఖ్యమైన పరికరం $f$-పీఠిక.

**నిర్వచనం 2.4.** ఒక $n$ కోష్టాలు గల $f$-**పీఠిక** అనే వస్తువుకు $కో_1, \ldots, కో_n$ అనే కోష్టాలుంటాయి. పైగ, క్రింది ఉపవిధులను భరిస్తుంది:

1. పీఠికలోని ప్రతి కోష్టము పఠ, లిఖ, పోల్చిమార్పు అనే నియోజ్యాలను సాధారణరీతిలో భరిస్తుంది

2. ప్రతి **ప** అనే $f$-పీఠిక, ఏదో ఒక ప్రత్యేకమైన నియోజ్యం $f$తో స్థాపింపబడి ఉంటుంది (ఉదా. $f$ = సంపర్కం, అనగా కూడిక). వస్తువుమీద మీద **ప**.$f()$ ని పిలిస్తే, అది $f(కో_1, \ldots, కో_n)$ను తిరుగిస్తుంది.

లేఖలో $f$-పీఠికలగురించిన ఒక ముఖ్యమైన సిద్ధాంతాన్ని వాడతాము.

**సిద్ధాంతం 2.5** ($f$-పీఠికలు [7]లో Theorem 6). $f$ అనే నియోజ్యం **లఘుతమ** (min), **బృహత్తమ** (max), లేదా **సంపర్క** (sum) ఐనచే, $f$-పీఠిక సమస్యకు ప్రతి సాధారణ నియోజ్యం (పఠ, లిఖ, పోల్చిమార్పు) $O(\log n)$ శ్రమతో, $f$-నియోజ్యమును $O(1)$ శ్రమతో పూర్తిచేసే విధికల్పము కలదు.

# 3 సామాన్య-జాగృతి-పరిష్కార సమస్య

ఈ భాగంలో జయంతి జాగృతి-పరిష్కారాన్ని సామాన్యపరుస్తాం. మా సామాన్య-జాగృతి-పరిష్కార సమస్య శ్రేణిలోని ప్రతి సమస్యను ఒక అప్రాసక (దుర్బల వర్ధన) శ్రేధి పరిమితి $s_1, \ldots, s_n$ ద్వారా విశదీకరిస్తాం. సామాన్య-జాగృతి-పరిష్కార సమస్య $జ(s_1, \ldots, s_n)$ అనగా, సుతథ్యముగా:

**నిర్వచనం 3.1** (జాగృతి-పరిష్కారం). ప్రతి వర్ధన శ్రేధి $0 \leq s_1 \leq s_2 \leq \cdots \leq s_n \leq n$ కి సంవాదిగా $జ(s_1, \ldots, s_n)$ అనే $n$ సంసాధక సంవిధానపు **సామాన్య-జాగృతి-పరిష్కార** సమస్య ఉంటుంది. ఆసమస్యను పూరించే విధికల్పమునకు మూడు గుణములుండాలి:

1. **సమాప్తి:** ప్రతి సంసాధకము $[1, n]$లో పూర్ణసంఖ్య తిరుగివ్వాలి.

2. **సత్యవాక్కు:** ఏ సంసాధకము $k$ సంసాధకాలు లేచేముందు, $k$ను తిరుగివ్వకూడదు.

3. **అవితండం:** $k$, లేదా యెక్కువ, సంసాధకాలు $s_k$ కంటే తక్కువ సంఖ్యను తిరుగివ్వకూడదు.

ఈ శ్రేణిలో $జ(1, 1, \ldots, 1, n)$ సమస్యను **సులభ జాగృతి-పరిష్కారం (సులభ సమస్య)** అని $జ(1, 2, \ldots, n-1, n)$ సమస్యను **కఠిన జాగృతి-పరిష్కారం (కఠిన సమస్య)** అని వ్యవహరిస్తాం.

సులభ జాగృతి-పరిష్కారము జయంతి జాగృతి-పరిష్కారానికి ప్రతిరూపం, యెందునంటే ఈ రెండు సమస్యలమధ్య సులభ-తరమైన పరస్పర న్యూనీకరణ సాధ్యము. వివరముగా, **వ** అనే విధికల్పము జయంతి జాగృతి-పరిష్కారాన్ని పూరిస్తే, **వ** ఎక్కడెక్కడ అసత్తును తిరుగిస్తుందో, అక్కడక్కడ 1ని, **వ** ఎక్కడెక్కడ సత్తును తిరుగిస్తుందో, అక్కడక్కడ $n$ను తిరుగిచ్చే **వ**′ అనే విధికల్పము సులభ జాగృతి-పరిష్కారాన్ని పూరిస్తుంది. ప్రతిదిశలో, **వ** అనే విధికపము సులభ జాగృతి-పరిష్కారాన్ని పూరిస్తే, **వ** ఎక్కడెక్కడైతే $n$ కంటే తక్కువ సంఖ్యను తిరుగిస్తుందో, అక్కడక్కడ అసత్తును, **వ** ఎక్కడెక్కడైతే $n$ను తిరుగిస్తుందో, అక్కడక్కడ సత్తును తిరుగిచ్చే **వ**′ సులభ జాగృతి-పరిష్కారాన్ని పూరిస్తుంది. అందుచేత, ఇప్పటినుంచి జయంతి జాగృతి-పరిష్కార సమస్యని కూడా సులభ సమస్యగానే వ్యవహరిస్తాం.

**సిద్ధాంతం 3.2** (సామాన్య జాగృతిపరిష్కార సిద్ధాంతం). $స_1, \ldots, స_n$ సంసాధకాల మధ్య సామాన్య జాగృతిపరిష్కార సమస్య $జ(s_1, \ldots, s_n)$ ని **వ** అనే వినా-ప్రతీక్ష విధికల్పం పూరిస్తుందని భావిద్దాం. అప్పుడు నిశ్చయముగా, ఏదోఒక కాలబంధం **క** ప్రకారం **వ** చలిస్తే, చలనకు $\Omega(n + \sum_{i=1}^{n} \log s_i)$ సంశ్రమ ఖర్చౌతుంది.

*నిరూపణ.* $జ(s_1, \ldots, s_n)$ ని **వ** అనే వినా-ప్రతీక్ష విధికల్పం పూరిస్తుందని భావిద్దాం. సిద్ధాంతం 2.2 లో ప్రస్తుతింపబడ్డ **క** అనే కాలబంధాన్ని పరీక్షిద్దాం. ఈ కాలబంధ క్రమంలో చలిస్తే ఏ **స** అనే సంసాధకము **త** అనే సంఖ్యను $\log_4 త$ కల్పల ముందు తిరుగివ్వలేదని విరోధోక్తి న్యాయం ద్వారా నిరూపిద్దాం.

స సంసాధకము క కాలబంధంలో అన్ని సంసాధకాలు చలిస్తున్నప్పుడు $త < \log_4 త$ అనే పూర్వ కల్పనలోనే త ను తిరుగించిందని భావిద్దాం. అవిచక్షణీయత ప్రకారం జ్ఞానగణంలో ఉన్న, అనగా జ్ఞా$_త$(స) లో ఉన్న, సంసాధకాలు మాత్రం క క్రమములో చలించినచో స సంసాధకము త ను త కల్పంలో తిరుగిస్తుంది. కానీ, సిద్ధాంతం 2.2 నలగవ అంశం ప్రకారం |జ్ఞా$_త$(స)| $\leq 4^త < 4^{\log_4 త} = త$, అందుచేత ఈ తిరిగిచ్చిన సంఖ్య సత్యవాక్యాన్ని వ్యతిరేకిస్తుంది. విరోధోక్తి సంపూర్ణం. తద్బ్రమాణముగ, త లేదా ఎక్కువ సంఖ్యను తిరుగిచ్చిన ప్రతి సంసాధకము కనీసం $\log_4 త$ శ్రమిస్తుందని గుర్తుంచుకుందాం.

అవితండము వలన $జ(s_1, \ldots, s_n)$ ని పూరించే ప్రతి విధికల్పంలో కనీసం ఒక సంసాధకము $s_n$ కి తక్కువకాని సంఖ్యను, కనీసం ఇంకొక సంసాధకము $s_{n-1}$ కి తక్కువకాని సంఖ్యను, ఇత్యాది. తిరిగివ్వాలి కనక, సంశ్రమ $\geq \sum_{i=1}^{n} \log_4 s_i = \Omega(\sum_{i=1}^{n} \log s_i)$ అని తేలింది. □

**ఉపసిద్ధాంతం 3.1** (కఠిన సమస్యాధోబంధం). $n$ సంసాధకాల సంవిధానంలో కఠిన సమస్యను పూరించే ప్రతి విధికల్పము ప్రతికూలతమ చలనలో $\Omega(n \log n)$ సంశ్రమిస్తుంది.

**ఉపసిద్ధాంతం 3.2** (సులభ సమస్యాధోబంధం). $n$ సంసాధకాల సంవిధానంలో సులభ సమస్యను పూరించే ప్రతి విధికల్పము ప్రతికూలతమ చలనలో $\Omega(n)$ సంశ్రమిస్తుంది.

పై నిరూపించిన అధోబంధాలను ఆసరాగా తీసుకొని న్యూనీకరణలద్వారా కొన్ని చేక్కటి క్రొత్త ఫలాలను నిరూపిస్తాం.

## 4 అధోబంధాలు

సమకాలిక కృత్రిమాణు దత్తాంశసంవిధానాల శ్రమవిషమతను తగ్గించే దిశలో ఈ మధ్య చాలా పని జరిగింది. ఈ ప్రగతిలో రెండు కీలకోదాహరణలు జయంతియొక్క గణత్రము [7], ఎలెన్ వోల్వ్ల్ల తెచ్చిమార్పు వస్తువు [3]. ఇరువస్తువులు ప్రతికూలతమశ్రమల్లో సర్వోత్తృష్టులని జయంతి అధోబంధాన్నిబట్టి తేలుతున్నది [6]. కానీ, సర్వసాధారణముగ ఒక్కటే వస్తు ఉపవిధిని పిలవడంకన్న చాలా వస్తూపవిధులను పిలవడం సంభవిస్తుంది. అందుకని ఒక్క ఉపవిధియొక్క ప్రతికూలతమశ్రమతో పాటు (ప్రతికూలతమ) సంశ్రమను గణించేడము ముఖ్యము. పై పేర్కొన్న రెండు వస్తువులు సంశ్రమవిషమతలో సర్వోత్తృష్టులని మా సామాన్య జాగ్రుతపరిష్కార అధోబంధం ద్వారా మొదట్టిసారిగా ఈ లేఖలో ప్రదర్శిస్తాం.

పరిశోధనలో అల్పవిషమత గల దత్తాంశసంవిధానాలను తయారు చేయలేని పక్షాలలో, సమీప-దత్తాంశసంవిధానాలు (తక్కువ ఆవశ్యకతలు పాటించే దత్తాంశసంవిధానాలు) సృజింపబడ్డాయి. ఉదాహరణకు, $h$-సమీప ప్రాధాన్యతాపంక్తి అనగా ప్రాధాన్యతాపంక్తిలోని న్యూనతమనిష్కాస ఉపవిధిలో, న్యూనతమ ప్రాధాన్యతగల సంఖ్యేగాక $h$ న్యూనతమ ప్రాధాన్యతలుగల సంఖ్యలలో ఏదైనా తిరుగివ్వచ్చు. ఆధునికముగా, న్యూనతమనిష్కాస ని $O(\log^3 n)$ అపేక్షిత అడుగుల్లో నిర్వహించే స్పేలిస్ట్ అనే $O(n \log^3 n)$-సమీప ప్రాధాన్యతాపంక్తిని ఆలిస్టా ఇత్యాదులు కల్పించారు [1, 2]. ఈ లేఖాభాగంలో అటువంటి సమీప-దత్తాంశసంవిధానాలనుదేశించే అధోబంధాలను కూడా ప్రదర్శిస్తాం.

సమీప-దత్తాంశసంవిధానలగురించిన సిద్ధాంతాలను వ్యక్తీకరించేడానికి క్రింది పరిభాషానిర్వచనాలు ఉపకరిస్తాయి.

**నిర్వచనం 4.1** (సమీప దత్తాంశసంవిధానాలు).

1. $h$-**సమీప గణత్రం**కి సామాన్య గణత్రముహోలేనే 0-విలువతో మొదలుగొని పెంచే(), విలువ() అనే ఉపవిధులను భరిస్తుంది. కానీ, విలువను తిరిగిచ్చినపుడు అసలు విలువకు $h$ వరకు యెక్కువైనా, తక్కువైనా గల సంఖ్యను తిరిగివ్వచ్చే.

2. $h$-**సమీప పంక్తి, రాశి, ప్రాధాన్యతాపంక్తి**లు ఆ ఆ సామాన్య వస్తువులుహోలే ఉంటాయి. కానీ, క్రమముల మొదటి సంఖ్యను నిశ్చయముగ తిరిగివ్వకుండ, క్రమముల మొదటి $h$ సంఖ్యలలో ఒకటి తిరుగిస్తాయి.

**సిద్ధాంతం 4.2** (గణత్ర అధోబంధాలు). గణత్రసదృశ వస్తువులు గురించిన కొన్ని సత్యాలు క్రింద ఇవ్వబడ్డాయి:

*1.* **గ** అనే విధికల్పము $n$ సంసాధకాలకు కృత్రిమాణు తెచ్చిపెంచు వస్తువని భావిద్దాం. ప్రతి $m \geq n$ కి ఏదో ఒక $m$ ఉపవిధుల అనుక్రమము $\Omega(m \log n)$ సంశ్రమను (ప్రతికూలతమములో) కలిగిస్తుంది.

2. **గ** అనే విధికల్పము $n$ సంసాధకాలకు కృత్రిమాణు గణత్రమని భావిద్దాం. ప్రతి $m \geq n$కి ఏదో ఒక $2m$ ఉపవిధుల అనుక్రమ-ము $\Omega(m \log n)$ సంశ్రమను (ప్రతికూలతములో) కలిగిస్తుంది.

3. **గ** అనే విధికల్పము $n$ సంసాధకాలకు కృత్రిమాణు $(1-\varepsilon)n/2$-సమీప గణత్రమని భావిద్దాం; యత్ర $\varepsilon > 0$ ఐన అచలం. ప్రతి $m \geq n$కి ఏదో ఒక $2m$ ఉపవిధుల అనుక్రమము $\Omega(m \log n)$ సంశ్రమను (ప్రతికూలతములో) కలిగిస్తుంది.

నిరూపణ. వాక్యాలను క్రమముగా నిరూపిస్తాం.

1. **గ** తొలి స్థితిలో 1 విలువతో ఉంటుంది. $m = n$ ఐతే, ప్రతి సంసాధకము తెచ్చిపెంచే() ను పిలిచి, వచ్చిన తిరుగువిలువను తిరుగిస్తే, కఠిన సమస్యకు పూరణొతుంది. అందున, నిరూపణ సిద్ధాంతం 3.1 ద్వారా సమాప్తము. వేరిక పూర్ణసంఖ్య $k$కు $m = kn$ ఐతే, పై ఉపాయమును కల్పాలలో పునస్కరించచ్చు. అనగా, కల్పాచివరల తడులు పెట్టి ఆగుతూ, ప్రతి కల్పములో కఠిన సమస్యను పూరిస్తే $k$ మాట్లు కఠిన సమస్యను పూరించినందుకు $\Omega(kn \log n) = \Omega(m \log n)$ సంశ్రమ తధ్యం. ఒకటే జాగ్రత్తేమిటంటే, $h^\text{వ}$ మాటు పున్స్కరించేనప్పుడు విలువనుంచి $(h-1)n$ను తీసివేసి $[1,n]$లో సంఖ్యను తిరిగివ్వాలి. ($m$ అనే సంఖ్య $n$ అనే సంఖ్యచేత భాగింపబడకపోతే, $n \times$ లబ్ధం మాట్లు పై మార్గమును అనుసరించి, శేషము మాట్లు ఉట్టినే ఉపవిధిని పిలిస్తే సరిపోతుంది.)

2. ప్రతి సంసాధకము **గ**.పెంచే()ని పిలిచి, తరువాత **గ**.విలువ()ను తిరుగిస్తే కఠిన సమస్య పూరింపబడుతుంది. (ఈ విషయా-న్ని విధికల్పం 5.2 ని ప్రస్తుతించినప్పుడు వివరిస్తాం.) అందుచేత $m = n$ ఐతే నిరూపణ సిద్ధాంతం 3.1 ద్వారా సమాప్తము. $k$ అనే పూర్ణ సంఖ్యకు $m = kn$ ఐతే, పై యుక్తిని $k$ మాట్లు, మధ్యన తడులు పెట్టి ఆగుతూ, అవలంబిస్తే సరిపోతుంది. ఒకటే జాగ్రత్తేమిటంటే, $h^\text{వ}$ మాటు అవలంబించేనప్పుడు విలువనుంచి $(h-1)n$ను తీసివేసి $[1,n]$లో సంఖ్యను తిరిగివ్వాలి.

3. ముందు $m = n$ అని భావించి పరిశీలిద్దాం. సమీప-గణత్రమునకు కూడా విధికల్పములో మొదట్ట **గ**.పెంచే()ని పిలవడం, తరువాత **గ**.విలువ()ను పిలవడం ఉంటాయి. కానీ వచ్చిన సంఖ్యను సరాసరి తిరుగివ్వకుండా, రాబోయే మార్పు చేస్తాం. తిరిగొచ్చిన విలువ **త** ఐనచో, తప్పకుండా కనీసం, **త** $-(1-\varepsilon)n/2$ సంసాధకాలు లేచాయని నిర్ధారించేచ్చు. అందున, ఆ సంఖ్యను (ఆ సంఖ్య ఒకటికంటే తక్కువైతే 1ని) తిరుగిస్తాం. మొదటి $n/2 + (1-\varepsilon)n/2 = (1-\varepsilon/2)n$ సంసాధకాలు **గ**ను పెంచిన తరువాత, **గ** యొక్క అసలు విలువ $(1-\varepsilon/2)n$ను దాటి ఉంటుంది. అందుకని చివరి $\varepsilon n/2$ సంసాధకాలు సమీప-గణత్రమైననూ $n/2$ను మించిన విలువలను చూసి, $n/2 - (1-\varepsilon)n/2 = \varepsilon n/2$ ను మించిన విలువలను తిరు-గిస్తాయి. అందుకని, ఈ వివరించిన విధికల్పము $s_1 = 1, \ldots, s_{n-\varepsilon n/2} = 1, s_{n-\varepsilon n/2+1} = \frac{\varepsilon}{2}n, \ldots, s_n = \frac{\varepsilon}{2}n$ పరిమితితో గల **జ**$(s_1, \ldots, s_n)$ సమస్యను పూరిస్తుంది. అందుచేత $\sum_{i=1}^n \log s_i = \varepsilon n/2 \log(\varepsilon n/2) = \Omega(n \log n) = \Omega(m \log n)$ సంశ్రమ వ్యయపరుస్తుంది (ప్రతికూలతమ చలనలో).

$m = kn$ ఐనచో, కల్పాలలో పునస్కరిస్తే సరిపోతుంది.

□

**ఉపవాక్యం.** సిద్ధాంతం 4.2 ప్రకారం జయంతి గణత్రము [7], ఎలెన్ వోల్పల్ల తెచ్చిపెంచే వస్తువు [3] సంశ్రమవిషమతలో సర్వోత్కృ-ష్టులు.

**సిద్ధాంతం 4.3.** **వ** అనే విధికల్పము కృత్రిమాణు $h$-సమీప పంక్తి కాని, రాశి కాని, ప్రాధాన్యతాపంక్తి కాని కానివ్వండి; యత్ర $h = (1-\varepsilon)n$ మరియు $\varepsilon > 0$ ఐన అచలం. ప్రతి $m \geq n$కి, ఏదో ఒక కాలబంధములో $m$ నిష్కాసనోపవిధుల అనుక్రమము $\Omega(m \log n)$ సంశ్రమను వ్యయపరుస్తుంది.

నిరూపణ. ముందు **వ** అనే వస్తువు పంక్తని భావిద్దాం; అలాగే $m = n$ అని భావిద్దాం. తొలి స్థితిలో **వ**లో క్రమముగా 1, 2, ..., $n$ ని ఉంచేదాం. అప్పుడు ప్రతి సంసాధకము ఒక మాటు నిష్కాసిస్తే, దానికి ఏదో ఒక సంఖ్య లభిస్తుంది. కనీసం $\frac{\varepsilon}{2}n$ సంసాధకా-లు నిష్కాసించే దాక $h$-సమీపత ప్రకారం, ఏ సంసాధకానికి $\frac{1-\varepsilon}{2}n$ ను మించిన సంఖ్య తిరుగదు. అలాగే ప్రతి సంసాధకానికి ఒక సంఖ్య వస్తుంది కాబట్టి, కనీసం $\frac{\varepsilon}{2}n$ సంసాధకాలకు $\frac{1-\varepsilon}{2}n$ ను మించిన సంఖ్య వస్తుంది. అందుచేత, నిష్కాసించినప్పుడు $\frac{1-\varepsilon}{2}n$ ను మించిన తిరుగుసంఖ్య వచ్చిన ప్రతి సంసాధకము $\frac{\varepsilon}{2}n$ను తిరుగిచ్చి, ఇతర సంసాధకాలన్నియు 1ని తిరుగిచ్చే విధికల్పము $s_1 = 1, \ldots, s_{n-\varepsilon n/2} = 1, s_{n-\varepsilon n/2+1} = \frac{\varepsilon}{2}n, \ldots, s_n = \frac{\varepsilon}{2}n$ పరిమితితో గల **జ**$(s_1, \ldots, s_n)$ సమస్యను పూరిస్తుంది. అందుచేత

$\sum_{i=1}^{n} \log s_i = \varepsilon n/2 \log(\varepsilon n/2) = \Omega(n \log n) = \Omega(m \log n)$ సంశ్రమ వ్యయపరుస్తుంది (ప్రతికూలతమ చలనలో). ఎప్పటిలాగే, $m > n$ ఐనచ, కల్పలలో పునస్కరిస్తే సరిపోతుంది.

ప్రాధాన్యతాపంక్తులకు ప్రాధాన్యతలు సంఖ్యలు ఒకటే అయ్యేడట్టు పెడితే ఇదే నిరూపణ సరిపోతుంది. రాశులకి క్రమమును తిరగేస్తే సరిపోతుంది. □

**ఉపవాక్యం.** పై ప్రదర్శించిన దత్తాంశసంవిధాన అధోబంధాలతో సైతం, ప్రస్తుతపు రచయిత జయంతి ఇత్యాద్యులు గోష్ఠీపత్రములో గణ-సంధి (set union) అనే వస్తువుకు కడు దుష్కరమైన న్యూనీకరణాధోబంధాన్ని స్థాపించేరు [8]. అలాగే, ఆ పత్రములో సూచింపబడ్డ కొన్ని సిద్ధాంతాలకు నిరూపణలు ఈ భాగంలో ఇప్పటికే విస్తృతింపబడి ఉన్నాయి.

## 5 ఉద్బంధాలు

### 5.1 సులభ సమస్య

ఈ లేఖలో మొట్టమొదటి క్రొత్త ఉద్బంధ-యోగదానముగా సులభ సమస్యకు అల్పసంశ్రమ విధికల్పమిస్తున్నాము. అవగమనార్థం, $m$ అనే పూర్ణసంఖ్యకు $n = 2^m$ అని భావిద్దాం. మా విధికల్పము (సంవిభక్తస్మృతిలో ఒక $n$ పర్ణలు గల $2n-1$ గ్రంథులు గల సమగ్ర ద్వయ వృక్షాన్ని స్థాపించి, ప్రతి గ్రంథిలో 0సంఖ్యను తొలిస్థితిలో వ్రాసి ఉంచేతుంది. ఆ పిమ్మట, స$_i$ సంసాధకము, క్రింది ఉపవిధిని $k = 1, x = i$వ పర్ణం అనే పరిమితులతో నిర్వహిస్తుంది:

---
**విధికల్పం 1** సరళ సమస్యను పూరించే విధికల్పములో వృక్షపవిధి.

    **ఉపవిధి** వృక్షపవిధి$(k, x)$
        **యదీ పోల్చిమార్పు**$(x, 0, k)$
            **యదీ** $x$ మూలం **తర్హీ తిరుగివ్వు** $2k$
            **అన్యథా యదీ** $x$.సోదర $\neq 0$ **తర్హీ తిరుగివ్వు** వృక్షపవిధి$(2k, x.$పితృ$)$
        **అన్యథా తిరుగివ్వు** $k$

---

**సిద్ధాంతం 5.1.** వృక్షపవిధి సరళ సమస్యను $O(n)$ సంశ్రమతో పూరిస్తుంది.

నిరూపణ. ముందు వృక్షపవిధి సరళ సమస్యను పూరిస్తుందని ప్రతిష్ఠిస్తాం. సరళ సమస్య సమాధానానికి మూడు గుణాలు. ప్రతి గుణాన్ని విడిగా నిరూపిస్తాం.

1. **సమాప్తి:** ప్రతి సంసాధకము దర్శించిన ప్రతి గ్రంథి వద్ద $O(1)$ శ్రమ మాత్రం ఖర్చుచేస్తుంది. ప్రతి సంసాధకము అధికతమముగా వృక్షోన్నతి, అనగా $\log n$, గ్రంథులని దర్శిస్తుంది కాబట్టి, సమాప్తి నిశ్చయం.

2. **సత్యవాక్కు:** ప్రతిష్ఠాపన ద్వారా $h$ ఉన్నతిలో ఉన్న గ గ్రంథిలో 0 లేకపోతే, ఆ గ్రంథిక్రిందున్న పర్ణాలలో మొదలైన $2^h$ సంసాధకాలు లేచాయని నిరూపిస్తాం. విధికల్పం తొలిస్థితిలో అన్ని గ్రంథుల్లో 0 ఉంటుందికాబట్టి ప్రతిష్ఠాపన మూలం నిలుస్తుంది. ప్రతి సంసాధకము స$_i$ వేరే పర్ణం ప$_i$తో మొదలిడుతుంది కాబట్టి, ప్రతి పర్ణం ప$_i$ దగ్గర ప్రతిష్ఠాపన నిలుస్తుంది. సంసాధకములు ఇరు పుత్రగ్రంథులు 0 కాదని స్థాపించుకున్న తరవాతే పితృ గ్రంథిని 0-కాని సంఖ్యకు పోల్చిమార్చడానికి ప్రయత్నిస్తాయి కాబట్టి ప్రతిష్ఠాపనపదం నిలుస్తుంది.

    సంసాధకము $h$ ఉన్నతికి చేరిన తరువాతే $k = 2^h$ తిరుగిస్తుంది కాబట్టి, సత్యవాక్కు నిశ్చయం.

3. **అవితండం:** ప్రతిష్ఠాపన ద్వారా ప్రతీ గ్రంథి $x$ని ఏదో ఒక సంసాధకం పోల్చిమార్చడానికి ప్రయత్నిస్తుందని నిరూపిస్తాం. ప్రతీ పర్ణము ఒక సంసాధక మూల స్థానము కాబట్టి ప్రతీ పర్ణమువద్ద ప్రతిష్ఠాపన నిలుస్తుంది. $x$ పర్ణేతర గ్రంథితే, దాని పుత్రగ్రంథులమీద పోల్చిమార్పులు జరిపిన సంసాధకాలలో చివర పోల్చిమార్పులో ఉత్తీర్ణమైన సంసాధకమును స$_i$ అని, పోల్చిమార్చిన పుత్రగ్రంథిని గలాని అందాం. ఆ సంసాధకము గ.సోదర $\neq 0$ అనే పరీక్ష చేసినప్పుడు సత్తు వస్తుంది. కావున ప్రతిష్ఠాపనపదం నిలుస్తుంది.

వృక్షమూలములు కూడా ఒక గ్రంథి కాబట్టి కనీసం ఒక సంసాధకమైన మూలాన్ని పోల్చిమార్చడానికి ప్రయత్నిస్తుంది. మూలాన్ని మార్చిన ప్రతి సంసాధకము $n$ ని తిరుగిస్తుంది. అందున అవితండం నిశ్చయం

శ్రమవిషమతను నిరూపించేడం జ్యామితిక శ్రేడీ సంపర్కము ద్వారా శులభం. ప్రతి గ్రంథిని ఏదో ఒక సంసాధకము మాత్రమును పోల్చిమార్చగలదు. పోల్చిమార్పును ప్రయత్నించి ఉత్తీర్ణమవని సంసాధకము వెంటనే తిరుగిస్తుంది. కాబట్టి, పోల్చిమార్పు ప్రయత్నాలు గ్రంథులసంఖ్యకు సంసాధకాలసంఖ్యను కలిపినకంటే ఉండలేవు. కావున, పోల్చిమార్పు ప్రయత్నాలు అధికతమముగ $2n-1+n=3n-1$. ఉపవిధిని పిలిచిన ప్రతిసారి పోల్చిమార్పు ప్రయత్నమౌతుంది కాబట్టి, సంశ్రమ $O(n)$. □

పై సిద్ధాంతాన్ని సులభ సమస్యాధీబంధంతో జోడించిన, వచ్చే ఉపసిద్ధాంతం మిగుల సులువుగా ఉత్పన్నమౌతుంది:

**ఉపసిద్ధాంతం 5.1.** *సులభ సమస్యయొక్క సంశ్రమ $\Theta(n)$.*

### 5.2 కఠిన సమస్య

సులభ సమస్యను సమూహఖేలన-వృక్షపవిధిద్వారా తక్కువ సంశ్రమతో పూరించేగలిగాము. ఇప్పుడు కఠిన సమస్య మీదకు దృష్టి సారిద్దాం. ఈ సమస్యకు ఇఠఃపూర్వ నిరూపించిన సిద్ధాంతప్రకారం అధీబంధం $\Omega(n\log n)$. ఈ సమస్యను సర్వోత్తమ సంశ్రమతో, అనగా $O(n\log n)$ సంశ్రమతో, పూరించచ్చేని $f$-పీఠికల ద్వార విధికల్ప ప్రమాణముగా నిరూపిస్తాం.

ఊహ సులభమోనదే. $f = $ సంపర్కం గా కల **గ** అనే $f$-పీఠికను పెట్టుకోని, క్రింది విధికల్పాన్ని ప్రతి సంసాధకము నిర్వహిస్తే చాలు:

---

**విధికల్పం 2** కఠిన సమస్యను పూరించే విధికల్పములో గణిత్రోపవిధి.

    **ఉపవిధి** గణిత్రోపవిధి()
        గ.పెంచే()
        **తిరుగివ్వు** గ.విలువ()

---

**సిద్ధాంతం 5.2.** *గణిత్రోపవిధి కఠిన సమస్యను $O(n\log n)$ సంశ్రమతో పూరిస్తుంది.*

*నిరూపణ.* **గ** కృత్రిమాణువస్తువు కనక సమాప్తి నిశ్చయం. **గ**ని ప్రతి సంసాధకము ఒక్కమాటే పెంచేతుంది కనక **గ** విలువ ఎప్పుడు లేచిన సంసాధక సంఖ్యను అధిగమించేదు; దాంతో సత్యవాక్కు నిశ్చయం. $k^{వ}$ స్థానములో మొదటి పంక్తిని పూర్తిచేసిన (రేఖింపబడ్డ) సంసాధకము కనీసం $k$ని తిరుగిస్తుంది; అందున అవితండము నిశ్చయం.

$f$-పీఠికల పూర్వాంశ సిద్ధాంత ప్రకారం ప్రతి సంసాధకము విషమతమ చలనలోకూడా $O(\log n + 1) = O(\log n)$ శ్రమిస్తుంది. అందున, సంశ్రమ $O(n\log n)$. □

పై నిరూపించేన ఉద్బంధము అధీబంధముతో కలుస్తుంది. అందుచేత క్రింది దృఢ పరిబంధం ఉత్పన్నమౌతుంది:

**ఉపసిద్ధాంతం 5.2.** *కఠిన సమస్యయొక్క సంశ్రమ $\Theta(n\log n)$.*

## 6 సమాప్తి

ఈ లేఖలో కృత్రిమాణుదత్తాంశసంవిధానాల సంశ్రమ విషమతను నిరూపించేటకు ఉపకరించే సామాన్య జాగృతి-పరిష్కారమనే సమస్యాశ్రేణిని నిర్వచించి పరిశీలించాము. పరిశీలనలో ముఖ్యభాగాలుమూడు, (1) శ్రేణిలోని ప్రతి సమస్యకు ఒక అధీబంధాన్ని నిరూపించేడం, (2) న్యూనీకరణల ద్వారా గణిత్రము, పంక్తి, రాశి, ప్రాధాన్యతాపంక్తి వంటి దత్తాంశసంవిధానాలకు అధీబంధాలను స్థాపించేడం, (3) జాగృతి-పరిష్కార కఠిన సమస్యకు ఉద్బంధమును నిరూపించి, ఆ ద్వారా ఈ పద్ధతికి గల పరిమితులను కూడా ప్రదర్శించేడం. మా జాగృతి పరిష్కార విధానము ద్వారా తొలి సారిగా, ఎలన్స్యల్లల తెచ్చిపెంచే దత్తాంశము [3], అలాగే జయంతి గణిత్రము [7] సంశ్రమ సర్వోత్తృష్టులని నిరూపించేగలిగాము. మా పద్ధతి ఈ ప్రస్తుత లేఖలోని సిద్ధాంతాలకగాక, జయంత్యాదుల

చేత మరింత దుష్కరమైన గణ-సంధి అధీబంధానికి [8] దారితీయడం గమనార్థం. గణ-సంధికి ఇంకా దృఢమైన అధీబంధాన్ని ఇవ్వడం (లేదా ఉద్బంధాన్ని దృఢపరచడం) కృత్రిమాణువస్తువులలో ఒక ముఖ్య ఉత్థాపిత సమస్యగా నిలచివుంది. జాగృతి-పరిష్కరణ ద్వారా ఇతర వస్తువులకు అధీబంధాలను నిరూపించేడము కూడా ఆకర్షణీయమైన దిశే.

శిక్షణాత్మకముగా చూసినచే, మా ఈ లేఖ తెలుగుభాషలో తొలి ఆధునిక సంగణక శాస్త్ర పరిశోధన పత్రికని మా నమ్మకము. కాని, సంస్కృతములో విలువైన గణిత, సంగణక శాస్త్ర సంపదగలదని ప్రసిద్ధము. ఆ సంప్రదాయమే తెలుగు శాస్త్రమునకు, ప్రత్యేకముగా ఈ లేఖకు స్ఫూర్తి. తదానుసారం, లేఖలోని పారిభాషికపదాలు చేలా మటుకు సంస్కృతరూపాంతరాలు; అందున సరళము-గా సంస్కృతములోకీ, ఇతర భారతీయభాషలలోకీ, అలాగే భారతేతరభాషలలోకీ అనువదింపశక్యముగా వుండునని మా ఊహ. తెలుగుభాషలోను, సంస్కృతభాషలోను నూతన ఆధునిక శాస్త్ర పురోగమన లేఖలు ఇత్యాదిగా కొనసాగాలని, ఆ లేఖల వలన ఎందరో స్ఫూర్తినీ శాస్త్ర జ్ఞానమును పొందగలరని, ఆత్మస్థైర్యాన్ని పెంపొందిచ్చేకొనగలరని మా అభిలాష.

## ఆధారాలు